\documentclass[amsmath,amssymb,aps,prd,10pt,twocolumn,showkeys]{revtex4}
\usepackage{graphicx}
\usepackage{mathtools}
\usepackage{verbatim}
\DeclareMathOperator\erfc{erfc}

\DeclareMathOperator{\sgn}{sgn}
\DeclareMathOperator{\snr}{SNR}
\begin{document}

\title{Complex Analysis of Askaryan Radiation: UHE-$\nu$ Identification and Reconstruction using the Hilbert Envelope of Observed Signals}

\author{Jordan C. Hanson}
\email{jhanson2@whittier.edu}
\affiliation{Department of Physics and Astronomy, Whittier College}
\author{Raymond Hartig}
\affiliation{Department of Physics and Astronomy, Whittier College}
\date{\today}

\begin{abstract}
The detection of ultra-high energy neutrinos (UHE-$\nu$), with enegies above 10 PeV, has been a long-time goal in astroparticle physics.  Autonomous, radio-frequency (RF) UHE-$\nu$ detetectors have been deployed in polar regions that rely on the Askaryan effect in ice for the neutrino signal.  The Askaryan effect occurs when the excess negative charge within a UHE-$\nu$ cascade radiates in a dense medium.  UHE-$\nu$ can induce cascades that radiate in the RF bandwidth above thermal backgrounds.  To identify UHE-$\nu$ signals in data from Askaryan-class detectors, analytic models of the Askaryan electromagnetic field have been created and matched to simulations and laboratory measurements.  These models describe the Askaryan electromagnetic field, but leave the effects of signal propagation through polar ice and RF channel response to simulations.  In this work, a fully analytic Askaryan model that accounts for these effects is presented.  First, formulas for the observed voltage trace and its Hilbert envelope are calculated.  Second, the analytic model is compared to UHE-$\nu$ signals at 100 PeV from NuRadioMC, a key Monte Carlo toolset in the field.  Correlation coefficients between the analytic signal envelope and MC data in excess of $0.94$ are found, and 99.99\% of UHE-$\nu$ signals pass a correlation threshold of $\rho\geq 0.4$.  Analysis of RF thermal noise reveals that just 0.2 background events have $\rho\geq 0.4$ in 5 years at a 1 Hz thermal trigger rate.  Finally, we describe future work related to the measurement of the logarithm of the UHE-$\nu$ cascade energy.
\end{abstract}


\maketitle

\section{Introduction}
\label{sec:int}

Cosmic neutrinos with energies up to 100 PeV have been detected by the IceCube and KM3NeT collaborations \cite{10.1126/science.1242856,aartsen2013first-bff,collaboration2016observation-03b,collaboration2018neutrino-2a0,collaboration2021detection-6fa,collaboration2022evidence-a08,collaboration2023observation-08b,collaboration2025observation-22f}. Previous analyses indicate that the discovery of UHE-$\nu$ flux above 5 PeV requires large Askaryan-class detectors \cite{10.1103/physrevd.98.062003}.  UHE-$\nu$ could reveal the source of ultra-high energy cosmic rays (see sections 3.1-3.3 of \cite{10.48550/arxiv.2008.04323}).  Further, studying electroweak interactions at these energies is impossible on Earth, and Askaryan-class neutrino detectors will provide new data (see section 3.4 of \cite{10.48550/arxiv.2008.04323}).

J. C. Hanson and R. Hartig presented the first fully analytic model of the Askaryan field in the time-domain (HH2022) \cite{PhysRevD.105.123019}.  The model builds on earlier work from J. C. Hanson and A. L. Connolly (JCH+AC), who developed an analytic form factor for the instantaneous charge distribution (ICD) that explains the low-pass filtering of the Askaryan spectrum, and cascade elongation from the LPM effect \cite{10.1016/j.astropartphys.2017.03.008}.  When correlated against semi-analytic parameterizations used in NuRadioMC (AHRZ2020), which involve numeric convolution of UHE-$\nu$ cascade data with an analytic vector potential, the HH2022 model yields correlation coefficients in excess of 0.95 \cite{PhysRevD.101.083005,PhysRevD.105.123019}.  Hanson and Hartig listed three advantages of analytic models in HH2022: measuring cascade parameters from the fit of analytic model to observed waveforms, minimization of computational intensity in NuRadioMC, and the potential to embed analytic models on-board detectors to reject thermal noise in real time.

Though the fully anaytic approach in HH2022 allowed precise reconstruction of UHE-$\nu$ cascade parameters, it was limited to comparisons between the simulated and theoretical $\vec{E}$-fields.  Askaryan-class detectors actually observe voltage traces that represent the RF detection channel response, convolved with $\vec{E}$-fields that have propagated through kilometers of polar ice.  NuRadioMC accounts for these effects by incorporating measurements from years of lab and field work \cite{10.1016/j.astropartphys.2014.09.002,10.3189/2015jog14j214,10.3189/2015jog15j057,saltzberg,10.1103/PhysRevD.74.043002,ask_ice,10.1140/epjc/s10052-020-7612-8,Barwick:2018497,ALLISON201963,10.1088/1748-0221/15/09/p09039,deaconu2018measurements-182,welling2024brief-b47}.  In this work, we present the first fully analytic Askaryan model in the time-domain that matches the observed voltage traces.

In practice, it is common to compute the Hilbert envelope of voltage traces from RF channels before cross-correlating them.  This is done to remove oscillations introduced by the RF antennas in the channels, which can confuse cross-channel timing and reconstruction.  Our calculations include both the voltage trace, and the Hilbert envelope of the trace.  The work is organized as follows.  Units, definitions, and notational conventions are given in Sec. \ref{sec:unit}.  The calculations of the observed voltage trace and Hilbert envelope of the trace are given in Sec. \ref{sec:onc}.  These results are compared to NuRadioMC output in Sec. \ref{sec:sim}.  In Sec. \ref{sec:recon}, a preliminary reconstruction of the logarithm of the cascade energy from the UHE-$\nu$ interaction is given.  The key findings are summarized in Sec. \ref{sec:conc}.

\section{Units, Definitions, and Conventions}
\label{sec:unit}

The analysis is based on two analytic functions: the Askaryan signal, $s(t)$, and the RF channel response, $r(t)$.  The RF detection channel is a linear, time-invariant DSP system.  Thus, the observed voltage trace in an Askaryan-class detector is the convolution of $r(t)$ and $s(t)$, $r(t) * s(t)$.  Let $\hat{s}(t)$ represent the Hilbert transform of $s(t)$, and let $j=\sqrt{-1}$.  The \textit{analytic signal}, $s_a(t)$, and \textit{signal envelope}, $\mathcal{E}_s(t)$, are defined by 

\begin{align}
s_a(t) &= s(t) + j\hat{s}(t) \\
\mathcal{E}_s(t) &= |s_a(t)|
\end{align}

The signal envelope actually observed by Askaryan-class detectors is the envelope of $r(t) * s(t)$, written as $\mathcal{E}_{r*s}(t)$.  The result for $\mathcal{E}_{r*s}(t)$ depends on the model for $s(t)$, taken to be Eq. 28 in HH2022 \cite{PhysRevD.105.123019}:
\begin{equation}
r\vec{E}(t_{\rm r},\theta) = -\frac{E_0\omega_0 \sin(\theta)}{8\pi p}t_{\rm r}e^{-\frac{t_{\rm r}^2}{4p}+p\omega_0^2}\erfc(\sqrt{p}\omega_0) \label{eq:s_full}
\end{equation}
\begin{table}
\begin{tabular}{| c | c | c |} \hline
\textbf{Variable} & \textbf{Definition} & \textbf{Units}\\ \hline
$c$ & speed of light in medium & m ns$^{-1}$ \\ 
$r$ & distance to cascade peak & m \\
$t_{\rm r}$ & $t-r/c$ & ns \\
$\theta_{\rm C}$ & Cherenkov angle & radians \\ 
$\theta$ & viewing angle from cascade axis & radians \\ 
$a$ & longitudinal cascade length (see \cite{10.1103/physrevd.65.016003}) & m \\ 
$n_{max}$ & max excess cascade particles (see \cite{10.1103/physrevd.65.016003})  & none \\
$E_{\rm 0}$ & $\propto n_{\rm max}a$ (see \cite{10.1103/physrevd.65.016003}) & V GHz$^{-2}$ \\
$p$ & $\frac{1}{2}(a/c)^2 \left(\cos\theta - \cos\theta_C\right)^2$ (see \cite{PhysRevD.105.123019}) & ns$^2$ \\ 
$\omega_{\rm 0}$ & $\sqrt{\frac{2}{3}} (c\sqrt{2\pi}\rho_{\rm 0})/(\sin\theta)$ (see \cite{10.1016/j.astropartphys.2017.03.008}) & GHz \\
$\sqrt{2\pi}\rho_{\rm 0}$ & lateral ICD width (see \cite{10.1016/j.astropartphys.2017.03.008}) & m$^{-1}$ \\ \hline
\end{tabular}
\caption{\label{tab:param} Parameters relevant for Eq. \ref{eq:s_full}.}
\end{table}

The parameters of Eq. \ref{eq:s_full} are shown in Tab. \ref{tab:param}.  Though Ralston and Buniy (RB) \cite{10.1103/physrevd.65.016003} used $c$ for the vacuum value of the speed of light, the formulae for $r \vec{E}$ presented in \cite{10.1103/physrevd.65.016003} refer to the wavenumber $k$ in the medium, which is proptional to the index of refraction. Thus, the use of $c$ in this work refers to the speed of light in the medium.  For example, a phase factor of $\exp(j k r)$ could also be written $\exp(j r \omega/c)$, if $c$ refers to the value in the medium.  The distance $r$ is between the observer and the radiating charge at the cascade peak.  The longitudinal length over which $\Delta r < \lambda$, the RF wavelength in ice, is named the \textit{coherence zone} $\Delta z_{\rm coh}$ in the RB model.  The $\Delta z_{\rm coh}$ is limited by what RB call the ``acceleration argument,'' that $r(t)$ is accelerating while keeping $\Delta r < \lambda$.

The time $t$ is the independent variable of the inverse Fourier transform of the equations in \cite{10.1103/physrevd.65.016003}. The delayed time is $t_{\rm r} = t-r/c$.  The speed $c$ is equal to the vacuum value, divided by the index of refraction $n$.  For RF in ice, $n=1.78$.  The viewing angle $\theta$ is the zenith angle in spherical coordinates, if the cascade axis is the z-axis.  The Cherenkov angle is $\cos\theta_{\rm C} = 1/n$ for relativistic cascades, and for the RF bandwidth in ice, $\theta_{\rm C} = 55.8$ degrees.  The longitudinal cascade length, $a$, is set by the cascade physics.  The ratio $\eta = (a/\Delta z_{\rm coh})^2$ corresponds to the far-field limit as $\eta \to 0$, but $\eta \to 0$ is not a requirement.  JCH+AC found that $\eta$ corresponds to a low-pass filter for the RF spectrum with cutoff frequency $\omega_{\rm C}$: $\eta = \omega/\omega_{\rm C}$ \cite{10.1016/j.astropartphys.2017.03.008}.

The $n_{\rm max}$ parameter is the maximum number of excess negative cascade charges, and the overall RF amplitude, $E_{\rm 0}$, is propotional to $n_{\rm max} a$ \cite{10.1103/physrevd.65.016003}.  JCH+AC and HH2022 demonstrated that the frequency $\omega_{\rm 0}$ is related to the ICD and the cascade form factor \cite{10.1016/j.astropartphys.2017.03.008,PhysRevD.105.123019}.  Monte Carlo simulations have shown that the lateral dependence of the ICD is exponentially distributed \cite{zhs,10.1016/j.astropartphys.2017.03.008}.  JCH+AC derived the form factor by modeling the lateral component of the ICD as an exponential distribution, which in turn makes the form factor act as a low-pass filter with cutoff $\omega_0$.  Finally, the authors of HH2022 showed that $p$ in Tab. \ref{tab:param} is related to $\sigma_t$, the pulse width of $s(t)$ \cite{PhysRevD.105.123019}:

\begin{equation}
\sigma_t = \sqrt{2p} \label{eq:pulse_width}
\end{equation}

The authors of HH2022 have shown that, because $\cos\theta - \cos\theta_{\rm C} \approx -\sin\theta_{\rm C}(\theta-\theta_{\rm C})$ to first order in $\Delta\theta=(\theta-\theta_{\rm C})$, $p \propto \Delta\theta^2$ to second order, and 

\begin{equation}
a\Delta\theta = \frac{c \sigma_t}{\sin\theta_{\rm C}} \label{eq:uncert}
\end{equation}

Qualitatively, this notion was identified by RB in Sec. III of \cite{10.1103/physrevd.65.016003}.  The authors of HH2022 analyzed the relationship between $a$, the cascade energy $E_{\rm C}$ and the critical energy $E_{\rm crit}$ for electromagnetic and hadronic cascades \cite{PhysRevD.105.123019}.  Let $E_{\rm C}/E_{\rm crit} = \Lambda$.  Assuming the Greisen and Gaisser-Hillas parameterizaions for electromagnetic and hadronic cascades, respectively, the following relationships for the $a$-values from electromagentically dominated and hadronically dominated cascades were found:

\begin{align}
a_{\rm em} &= x_{\rm em} \sqrt{\ln\Lambda} \label{eq:em} \\
a_{\rm had} &= x_{\rm had} \sqrt{\ln\Lambda} \label{eq:had}
\end{align}
\noindent
The values for $x_{\rm em}$ and $x_{\rm had}$ are calculated in Sec. \ref{sec:recon}. From Eq. \ref{eq:uncert}, the fractional error in $\ln\Lambda$ is:

\begin{equation}
\frac{\sigma_{\ln\Lambda}}{\ln\Lambda} = 2\left(\frac{\sigma_a}{a}\right) \label{eq:a_err}
\end{equation}
\noindent
Equation \ref{eq:a_err} corresponds to Eq. 42 in \cite{PhysRevD.105.123019}, and has been corrected for units.  Equations \ref{eq:pulse_width}-\ref{eq:a_err} imply measurements of $a$ and $\Delta\theta$ yield $\ln\Lambda$, and that the relative error in $\ln\Lambda$ is proportional to the relative error in $a$.

\section{Calculation of the Main Results}
\label{sec:onc}

The parameters in Eq. \ref{eq:s_full} that do not depend on time can be folded into a single constant, $E_0$, leaving only the time-dependence.  From now on, let $t$ refer to $t_{\rm r}$, without the subscript.  The signal model $s(t)$ is
\begin{equation}
s(t) = -E_0 t e^{-\frac{1}{2}\left(t/\sigma_t\right)^2} \label{eq:s}
\end{equation}
This is the \textit{off-cone} equation from \cite{PhysRevD.105.123019}.  The parameter $\sigma_{\rm t}$ is the pulse width, and it depends two quantities: $a$ and $\Delta\theta$ (Eq. \ref{eq:uncert}).  The parameter $E_0$ is the amplitude normalization, and the dependencies on other parameters can be determined from Eq. \ref{eq:s_full} and Tab. \ref{tab:param}.  The most important of these is the $1/r$ dependence (Eq. \ref{eq:s_full}).  To achieve the goal of $\mathcal{E}_{\rm r * s}(t)$, the Hilbert transform and analytic signal of $s(t)$ are required.  The Hilbert transform $\widehat{s}(t)$ is equivalent to the convolution of $s(t)$ and the tempered distribution $h(t) = 1/(\pi t)$.  The implication in the Fourier domain is that the negative frequencies in the spectrum of $\hat{s}(t)$ vanish, while the positive ones are doubled.  Let the $\sgn(f)$ be $-1$ if $f<0$, $0$ if $f=0$, and $1$ if $f>1$, and let $S(f)$ be the Fourier transform of $s(t)$.  The Fourier transform of the analytic signal $s_a(t)$ is
\begin{equation}
\mathcal{F}\lbrace s_{\rm a}(t) \rbrace_{f} = S_{\rm a}(f) = S(f)(1+\sgn{f}) \label{eq:sa_1}
\end{equation}
Thus, if $f<0$, $S_a(f) = 0$, and $S_a(f) = 2S(f)$ if $f\geq 0$.  Taking the inverse Fourier transform of Eq. \ref{eq:sa_1}, the analytic signal may be written in terms of $S(f)$:
\begin{equation}
s_{\rm a}(t) = 2\int_{0}^{\infty} S(f) e^{2\pi j f t} df \label{eq:sa_2}
\end{equation}
 The Fourier transform of Eq. \ref{eq:s} is
\begin{equation}
S(f) = E_0 \sigma_t^3 (2\pi)^{3/2} j f e^{-2\pi^2 f^2 \sigma_t^2}
\end{equation}
Using the gaussian spectral width $\sigma_{\rm f}$ from \cite{10.1016/j.astropartphys.2017.03.008}, and the guassian width of $s(t)$ from \cite{PhysRevD.105.123019}, it was shown in \cite{PhysRevD.105.123019} that the uncertainty principle holds for off-cone signals:
\begin{equation}
\sigma_t \sigma_f \geq \frac{1}{2\pi}
\end{equation}
The equality is reached in the limit the far-field parameter limits to zero: $\eta \to 0$.  This makes the signal spectrum
\begin{equation}
S(f) = E_0 \sigma_t^3 (2\pi)^{3/2} j f e^{-\frac{1}{2}\left(f/\sigma_f\right)^2} \label{eq:spec}
\end{equation}
Inserting $S(f)$ into Eq. \ref{eq:sa_2}, $s_{\rm a}(t)$ is
\begin{equation}
s_{\rm a}(t) = \frac{E_0 \sigma_t^3 (2\pi)^{3/2}}{\pi} \frac{d}{dt}\int_0^{\infty} e^{-\frac{1}{2}\left(f/\sigma_f\right)^2} e^{2\pi j f t} df \label{eq:sa_3}
\end{equation}
 Let $k^2/4 = \frac{1}{2}\left(f/\sigma_f\right)^2$, and $x = t/(\sqrt{2}\sigma_t)$.  Equation \ref{eq:sa_3} can be broken into real and imaginary parts:
\begin{align}
s_{\rm a}(t) &= \frac{E_0 \sigma_{\rm t}}{\sqrt{2\pi}}\frac{dI}{dx} \\
\Re\lbrace I \rbrace &= \int_0^{\infty} e^{-k^2/4}\cos(kx) dk \\
\Im\lbrace I \rbrace &= \int_0^{\infty} e^{-k^2/4}\sin(kx) dk
\end{align}
The real part of $I$ is even, so it can be extended to $(-\infty,\infty)$ if it is multiplied by $1/2$.  The result is
\begin{equation}
\Re\lbrace I \rbrace = \sqrt{\pi} e^{-x^2}
\end{equation}
The imaginary part of $I$ is proportional to the \textit{Dawson function, $D(x)$} \cite{NIST:DLMF}:
\begin{equation}
\Im\lbrace I\rbrace = 2 D(x)
\end{equation}
 The overall analytic signal, $s_a(t)$, is
\begin{equation}
s_a(t) = -E_0 \left(t e^{-\frac{1}{2}\left(t/\sigma_t\right)^2} - \frac{2 j\sigma_t}{\sqrt{2\pi}} \frac{dD(x)}{dx}\right) \label{eq:sa_4}
\end{equation}
The envelope of the signal, $\mathcal{E}_s(t)$, is the magnitude of Eq. \ref{eq:sa_4}.  Though $D(x)$ is not evaluated analytically, a high-precision algorithm for computing $D(x)$ was given in \cite{10.1063/1.4822832}.  As expected, $|s_a(0)| \neq 0$, since $dD(x)/dx = 1 - 2x D(x)$.

The observed data in Askaryan-class detectors is the convolution of the UHE-$\nu$ signal from the ice and the detector response function.  To generate \textit{signal templates}, Askaryan radiation signals are calculated for the UHE-$\nu$ interaction properties, modified by the frequency-dependent RF attenuation of polar ice, and convolved with the RF channel response \cite{10.1016/j.astropartphys.2014.09.002,10.3189/2015jog14j214}.  Signal templates are cross-correlated with observed data to identify UHE-$\nu$ signals.  RF detection channels based on RF dipole antennas, however, have resonance frequencies that introduce oscillations not present in the original signal.  The oscillations can introduce timing uncertainties.  The problem intensifies when the signal-to-noise ratio (SNR) relative to RF thermal noise decreases.

To reduce uncertainties, the Hilbert envelope of observed data is used in cross-correlations instead of the original signals.  Thus, an analytic prediction for the Hilbert envelope of the observed data would an effective signal template.  An assumption must be made, however, for the RF channel response, $r(t)$.  The RLC impulse response was first used by RICE at the South Pole at a model for RF dipole channels \cite{10.1103/PhysRevD.85.062004}.  The RLC damped oscillator is a suitable circuit model for the RF dipole channels for two reasons.  First, dipoles respond within a bandwidth centered around a resonance frequency, just as RLC circuits do.  Second, RF dipole channels are used in RICE, ARA, RNO-G, and the proposed IceCube Gen2 because these channels must fit inside cylindrical ice boreholes \cite{10.1103/PhysRevD.85.062004,10.1103/physrevd.102.043021,allison2012design-cd3,10.1088/1748-0221/16/03/p03025,10.48550/arxiv.2008.04323}.  Thus, the model chosen for $r(t)$ is the damped oscillation of an RLC circuit.

There are two paths to calculating the final result, $\mathcal{E}_{r*s}(t)$.  The first involves three steps.  First, the detector response, $r(t)$ is convolved with $s(t)$.  Second, the analytic signal of the result is found.  Third, the magnitude of the analytic signal is computed, which can be compared to envelopes of observed signals.  The second path involves computing $\mathcal{E}_{r*s}(t)$ directly from $s_a(t)$ and $r_a(t)$.  The second option is more straightforward, once a special theorem relating $r_a(t)$, $s_a(t)$, and $\mathcal{E}_{r*s}(t)$ is established.  $\mathcal{E}_{s * r}(t)$, $s_a(t)$, and $r_a(t)$ are related by
\begin{equation}
\mathcal{E}_{s * r}(t) = \frac{1}{2}| s_a (t) * r_a(t)| \label{eq:awesome}
\end{equation}

The proof of Eq. \ref{eq:awesome} is based on two ideas.  First, the Hilbert transform of a function $s(t)$ is equivalent to convolving it with $h(t) = 1/(\pi t)$.  Second, the Hilbert transform is an \textit{anti-involution}, meaning acting it twice on a function $f(t)$ yields $-f(t)$: $h * h * f = -f$.  Given the definitions of $s_a(t)$ and $\hat{s}(t)$,
\begin{align}
(s * r)_a (t) &= s * r + j ~ \widehat{s*r} \\
\mathcal{E}_{s * r}(t) &= | s * r + j s * r * h|
\end{align}
However,
\begin{align}
r_a * s_a &= (r + j \hat{r}) * (s + j \hat{s}) \\
r_a * s_a &= r * s + j r * \hat{s} + j \hat{r} * s - \hat{r} * \hat{s} \\
r_a * s_a &= r * s - r * h * s * h + 2 j h * r * s \\
r_a * s_a &= r * s - h * h * r * s + 2 j h * r * s \\
r_a * s_a &= 2 r * s + 2 j h * r * s
\end{align}
Multiplying both sides $1/2$ and taking the magnitude completes the proof:
\begin{equation}
\frac{1}{2} |r_a * s_a| = |r * s + j h * r * s| = \mathcal{E}_{s * r}(t) \\
\end{equation}

Assume that a signal arrives in an RLC damped oscillator at $t=0$.  For $t\geq 0$, the impulse response and corresponding analytic signal are
\begin{align}
r(t) &= R_0 e^{-2 \pi \gamma t} \cos(2\pi f_0 t) \label{eq:r} \\
r_a(t) &= R_0 e^{-2 \pi \gamma t} e^{2\pi j f_0 t} \label{eq:ra}
\end{align}
The parameters $\gamma$ and $f_0$ are the decay constant and the resonance frequency.  Note that the envelope of $r(t)$, $|r_a(t)|$, is $R_0 \exp(-2 \pi \gamma t)$, as expected.  To prove Eq. \ref{eq:ra}, first compute the Fourier transform of $r(t)$:
\begin{align}
R(f) &= \frac{R_0}{4\pi j} \left( \frac{1}{f - z_{+}} + \frac{1}{1- z_{-}} \right) \\
z_{+} &= f_0 + j \gamma \\
z_{-} &= -f_0 + j \gamma
\end{align}
Given Eq. \ref{eq:sa_2}, the procedure to find $r_a(t)$ is to multiply the \textit{negative} frequency component at $z_{-}$ by 0 and the \textit{positive} frequency component at $z_{+}$ by 2, and take the inverse Fourier transform.  The inverse Fourier transform may be completed by applying Jordan's lemma in the complex plane frequency plane.  The residue from the pole at $z_{+}$ yields the result.

The goal is now to apply Eq. \ref{eq:awesome} by convolving $s_a(t)$ with $r_a(t)$.  The calculation may be split into two parts: $r_a(t) * \Re\lbrace s_a(t) \rbrace$, and $r_a(t) * \Im\lbrace s_a(t) \rbrace$.  Let $u(t)$ represent the Heaviside step function.  Starting with $r_a(t) * \Re\lbrace s_a(t) \rbrace$:
\begin{multline}
r_a(t) * \Re\lbrace s_a(t) \rbrace = \\ R_0 e^{2\pi j f_0 t} e^{-2\pi\gamma t} u(t) * \left(-E_0 t e^{-\frac{1}{2}\left(t/\sigma_t\right)^2}\right)
\end{multline}
Let $x=t/(\sqrt{2}\sigma_t)$, $y=\tau/(\sqrt{2}\sigma_t)$, and $z = (2\pi j f_0 - 2\pi\gamma)\sqrt{2}\sigma_t$.  Changing variables while accounting for the relationship between $u(t)$, $x$, and $y$, gives
\begin{multline}
r_a(t) * \Re\lbrace s_a(t) \rbrace = \\ -2R_0 E_0 \sigma_t^2 \int_{-\infty}^{x} e^{z(x-y)} y e^{-y^2} dy
\end{multline}
Note that the units for the convolution of $r(t)$ and $s(t)$ are $R_0 E_0 \sigma_t^2$.  Let $u = x-y$, so that $du = -dy$. The result is
\begin{equation}
r_a(t) * \Re\lbrace s_a(t) \rbrace = 2R_0 E_0 \sigma_t^2\left(\frac{dI(x,z)}{dz}-xI(x,z)\right)
\end{equation}
where
\begin{equation}
I(x,z) = \int_0^{\infty} e^{zu} e^{-(u-x)^2} du
\end{equation}
Let $b = x+\frac{1}{2} z$. Completing the square in the exponent and substituting $k = u-b$ gives
\begin{multline}
I(x,z) = e^{-x^2} e^{b^2} \int_{-b}^{\infty} e^{-k^2} dk \\ = \frac{\sqrt{\pi}}{2} e^{-x^2} e^{b^2} \erfc(-b)
\end{multline}
Let $b = jq$, and $w(q)$ be the \textit{Faddeeva function} \cite{NIST:DLMF}.  The integral becomes
\begin{equation}
I(x,z) = \frac{\sqrt{\pi}}{2} e^{-x^2} w(q)
\end{equation}
The chain rule is required to find $dI/dz$:
\begin{equation}
\frac{dI}{dz} = \frac{dI}{dq}\frac{dq}{dz} = -\left(\frac{j}{2}\right)\frac{dI}{dq}
\end{equation}
The final result is
\begin{multline}
r_a(t) * \Re\lbrace s_a(t) \rbrace = \\ -\sqrt{\pi} R_0 E_0 \sigma_t^2 \left(x e^{-x^2} w(q) + \left(\frac{j}{2}\right) e^{-x^2} \frac{dw(q)}{dq} \right) \label{eq:Re_result}
\end{multline}

Turning to the convolution of $r_a(t)$ with $\Im(s_a)$,
\begin{multline}
r_a(t) * \Im\lbrace s_a(t) \rbrace = \\ \left(R_0 e^{2\pi j f_0 t} e^{-2\pi \gamma t} u(t)\right) * \left(\frac{2 E_0 \sigma_t^2}{\sqrt{\pi}}\frac{dD(t/\sqrt{2}\sigma_t)}{dt} \right)
\end{multline}
Note that $f'(t) * g(t) = f(t) * g'(t) = (f(t) * g(t))'$.  Thus,
\begin{multline}
r_a(t) * \Im\lbrace s_a(t) \rbrace = \\ \frac{2}{\sqrt{\pi}} R_0 E_0 \sigma_t^2 \frac{d}{dt}\left(e^{2\pi j f_0 t}e^{-2\pi\gamma t} u(t) * D(t/\sqrt{2}\sigma_t) \right)
\end{multline}
The convolution becomes
\begin{multline}
r_a(t) * \Im\lbrace s_a(t) \rbrace = \\ \frac{2}{\sqrt{\pi}} R_0 E_0 \sigma_t^2 \frac{d}{dt} \int_{-\infty}^{t} e^{(2\pi j f_0 - 2\pi\gamma)(t-\tau)}D(\tau/\sqrt{2}\sigma_t)d\tau
\end{multline}
Adopting the earlier definitions of $x$, $y$, and $z$ gives
\begin{multline}
r_a(t) * \Im\lbrace s_a(t) \rbrace = \\ \frac{2}{\sqrt{\pi}} R_0 E_0 \sigma_t^2 \frac{d}{dx} \int_{-\infty}^{x} e^{z(x-y)} D(y) dy
\end{multline}
Using Leibniz rule for the fundamental theorem of calculus, and the limiting cases of $D(x)$,
\begin{multline}
r_a(t) * \Im\lbrace s_a(t) \rbrace = \\ \frac{2}{\sqrt{\pi}} R_0 E_0 \sigma_t^2 \left(D(x) + z\int_{-\infty}^{x} e^{z(x-y)} D(y) dy \right)
\end{multline}
Let $u = x-y$, $z=-k$, and note that $D(x)$ is an odd function.  These substitutions give
\begin{multline}
r_a(t) * \Im\lbrace s_a(t) \rbrace = \\ \frac{2}{\sqrt{\pi}} R_0 E_0 \sigma_t^2 \left(D(x) + k\int_{0}^{\infty} e^{-ku} D(u-x) du \right)
\end{multline}
The remaining integral is the Laplace transform of the shifted Dawson function, $\mathcal{L}\lbrace D(u-x)\rbrace_k$.  The final result is
\begin{multline}
r_a(t) * \Im\lbrace s_a(t) \rbrace = \\ \frac{2}{\sqrt{\pi}} R_0 E_0 \sigma_t^2 \left(D(x) + k\mathcal{L}\lbrace D(u-x)\rbrace_k\right) \label{eq:Im_result}
\end{multline}
Though a closed analytic form for $\mathcal{L}\lbrace D(u-x)\rbrace_k$ is elusive, evaluating $\mathcal{L}\lbrace D(u-x)\rbrace_k$ numerically is fast and precise. A short code that demonstrates how this Laplace transform is computed is given in Appendix \ref{sec:appA}.

Combining Eq. \ref{eq:Re_result} and Eq. \ref{eq:Im_result} gives $r_a(t) * s_a(t)$:
\begin{multline}
r_a(t) * s_a(t) = \\ -\sqrt{\pi} R_0 E_0 \sigma_t^2 \left(x e^{-x^2} w(q) + \left(\frac{j}{2}\right) e^{-x^2} \frac{dw(q)}{dq} \right) + \\ \frac{2j}{\sqrt{\pi}} R_0 E_0 \sigma_t^2 \left(D(x) + k\mathcal{L}\lbrace D(u-x)\rbrace_k\right) \label{eq:final}
\end{multline}
The units of convolution should be $R_0 E_0\sigma_t^2$, and each term in Eq. \ref{eq:final} has these units.  Remember that the relationship between $q$ and $x$ is given by
\begin{equation}
q = -jb = -j\left(x + \frac{z}{2}\right) \\
\end{equation}
Taking the magnitude of Eq. \ref{eq:final}, and multiplying by $1/2$, yields the \textbf{Hilbert envelope of the convolution of $s(t)$ with $r(t)$:}
\begin{equation}
\mathcal{E}_{r * s}(t) = \frac{1}{2} | r_a(t) * s_a(t) | \label{eq:final2}
\end{equation}

To model the full time-dependent voltage trace from detection channels, and not merely the envelope, $s(t) * r(t)$ is required.  The calculation may be done analytically, starting with the definition of convolution:
\begin{equation}
s(t) * r(t) = \int_{-\infty}^{\infty} s(t-\tau) r(\tau) d\tau 
\end{equation}
As in the calculation of $\mathcal{E}_{r*s}(t)$, the Heaviside step function, $u(t)$ is included to ensure causality.  Inserting the definitions of $s(t)$, $r(t)$, and $u(t)$ gives
\begin{multline}
s * r = \\ -E_0 R_0 \int_{-\infty}^{\infty} (t-\tau) e^{-\frac{1}{2}\left(\frac{t-\tau}{\sigma_t}\right)^2} \Re\left\lbrace e^{2\pi j f_0 \tau} e^{-2\pi\gamma \tau} \right\rbrace u(\tau) d\tau
\end{multline}
Using the previous definitions of $x$, $y$, and $z$ gives
\begin{equation}
s * r = -2R_0 E_0 \sigma_t^2 \int_{0}^{\infty} (x-y) e^{-(x-y)^2} \Re\left\lbrace e^{zy} \right\rbrace dy
\end{equation}
Note that the $\Re\lbrace \rbrace$ operator can encompass the whole integral, since $s(t)$ is real.  Splitting the integral and employing differentiation under the equals sign yields
\begin{multline}
s * r = \\ -2R_0 E_0 \sigma_t^2 \Re\left\lbrace xe^{-x^2}I(x,z) - \frac{1}{2}e^{-x^2} \frac{dI(x,z)}{dx} \right\rbrace \label{eq:conv_sr}
\end{multline}
with
\begin{equation}
I(x,z) = \int_0^{\infty} e^{-y^2 + (2x + z)y}dy
\end{equation}
Let $b=x+\frac{1}{2}z$, and $b = j q$.  The calculation resembles that of $r_a(t) * \Re\lbrace s_a(t)\rbrace$.  The result for $I(x,z)$ is
\begin{equation}
I(x,z) = \frac{\sqrt{\pi}}{2} w(q)
\end{equation}
Inserting this result into Eq. \ref{eq:conv_sr}, gives
\begin{multline}
s * r = -\sqrt{\pi}R_0 E_0 \sigma_t^2 \\ \Re\left\lbrace xe^{-x^2} w(q) - \frac{1}{2}e^{-x^2} \frac{d w(q)}{dx} \right\rbrace
\end{multline}
From the definition of $q$ and the chain rule, $dw(q)/dx = -jdw(q)/dq$, and $dw(q)/dq = -2qw(q)+2j/\sqrt{\pi}$ \cite{NIST:DLMF}.  The final result is left in terms of $\Re\lbrace w(q)\rbrace$ and $\Re\lbrace -jdw(q)/dq\rbrace$, which are proportional to the \textit{Voigt functions} \cite{NIST:DLMF,PhysRevD.105.123019}.
\begin{multline}
s * r = -\sqrt{\pi}R_0 E_0 \sigma_t^2 \\ \left(xe^{-x^2} \Re\left\lbrace w(q) \right\rbrace - \frac{1}{2}e^{-x^2} \Re\left\lbrace -j \frac{d w(q)}{dq} \right\rbrace \right) \label{eq:final3}
\end{multline}

\begin{figure}[ht]
\centering
\includegraphics[width=0.5\textwidth]{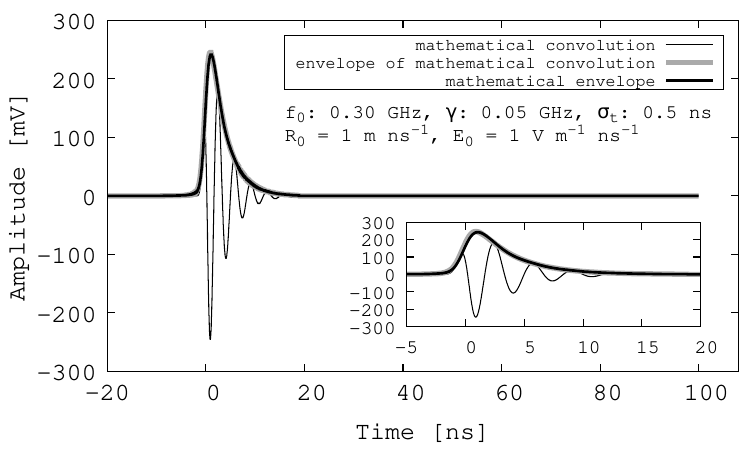}
\includegraphics[width=0.5\textwidth]{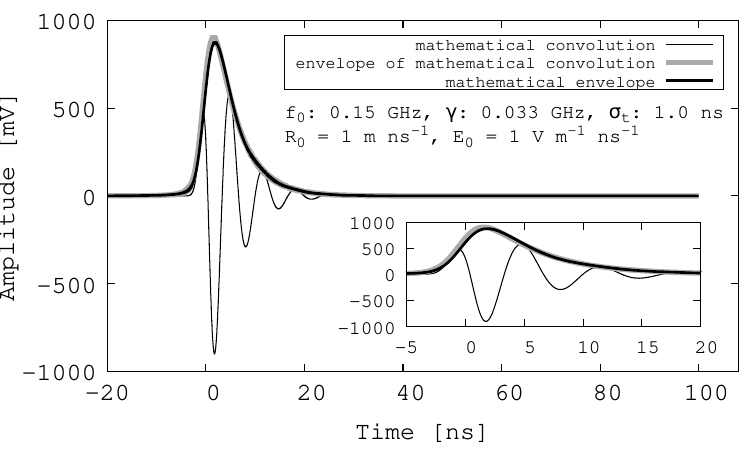}
\caption{\label{fig:fig1} (Top) The thin black line represents $s * r$ from Eq. \ref{eq:final3}.  The gray envelope represents the envelope of Eq. \ref{eq:final3} computed with the Python3 SciPy function scipy.special.hilbert. The black envelope represents $\mathcal{E}_{r * s}(t)$ from Eqs. \ref{eq:final}-\ref{eq:final2}. (Bottom) Same as top, different parameters.}
\end{figure}

\begin{figure}[ht]
\centering
\includegraphics[width=0.5\textwidth]{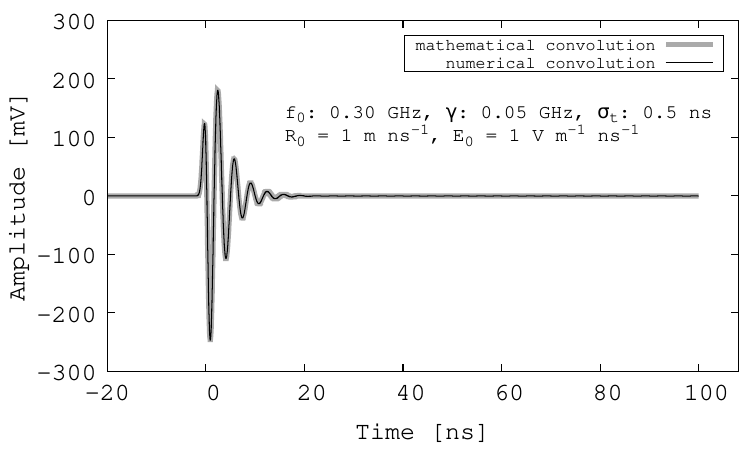}
\includegraphics[width=0.5\textwidth]{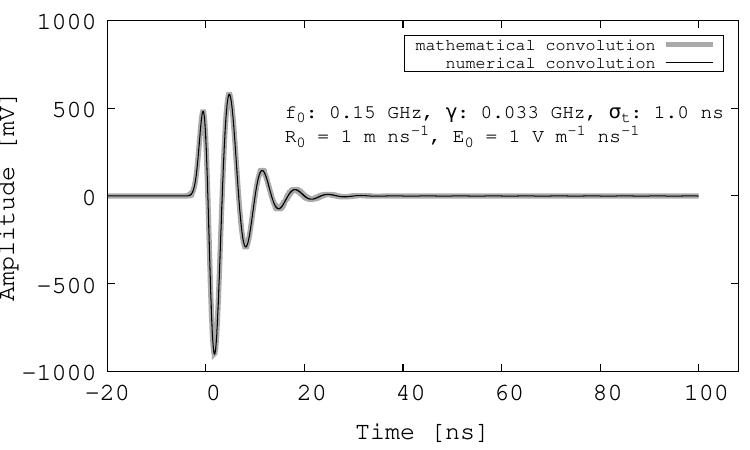}
\caption{\label{fig:fig2} (Top) The thin black line represents $s(t)$ from Eq. \ref{eq:s} convolved with $r(t)$ from Eq. \ref{eq:r}, using the Python3 SciPy function scipy.signal.convolve. The gray line represents $s * r$ from Eq. \ref{eq:final3}. (Bottom) Same as top, different parameters.}
\end{figure}

To illustrate the accuracy of the model, the \textit{numerically-computed envelope} of Eq. \ref{eq:final3} is compared to Eqs. \ref{eq:final}-\ref{eq:final2} in Fig. \ref{fig:fig1}.  The \textit{numerical convolution} of $s(t)$ and $r(t)$ is compared to Eq. \ref{eq:final3} in Fig. \ref{fig:fig2}.

\section{Comparison between Analytic Calculations and NuRadioMC}
\label{sec:sim}

NuRadioMC was used to generate UHE-$\nu$ interactions in a cylindrical ice volume (see Tab. \ref{tab:1}).  The 100 PeV interactions included charged and neutral electroweak currents, and the LPM effect.  The Askaryan model used in NuRadioMC was AHRZ2020 \cite{PhysRevD.101.083005}.  AHRZ2020 is a \textit{semi-analytic} parameterization of the Askaryan effect, in which a vector potential $\vec{A}(\vec{r},t)$ is convolved with the profile of the net charge in the UHE-$\nu$ cascade.  AHRZ2020 accounts for sub-cascades and the LPM effect, and it has been validated against MC simulations \cite{zhs,10.1103/physrevd.84.103003}.  The Askaryan model used in NuRadioMC for this analysis was \textit{not} HH2022, so correlations between MC output and Eqs. \ref{eq:final}-\ref{eq:final2} are solely on physical grounds.

The detector was \textit{a single string} of 8 RF dipoles, each with the same (x,y)-coordinates and regularly-spaced z-coordinates.  The ice had a depth and radius of 0.65 km and 0.85 km, respectively, and the accounted for measurements of the attenuation length versus frequency collected from the South Pole region \cite{Barwick_Besson_Gorham_Saltzberg_2005,allison2012design-cd3}.  Each RF channel had a filtered, amplified passband of [0.08-1] GHz, 256 samples, and a 1 GHz sampling rate.  The RF trigger responded when any 3 of the 8 voltage traces exceeded $+3$ times \textit{and} $-3$ times the rms voltage ($v_{\rm rms}$) of the thermal noise within 256 ns (256 samples).  The rms voltage, $v_{\rm rms}$, is characterized in NuRadioMC via the \textit{noise temperature}, which was set to 233 K.

\begin{figure}
\centering
\includegraphics[width=0.5\textwidth,trim=3.25cm 8.25cm 4.5cm 9.0cm,clip=true]{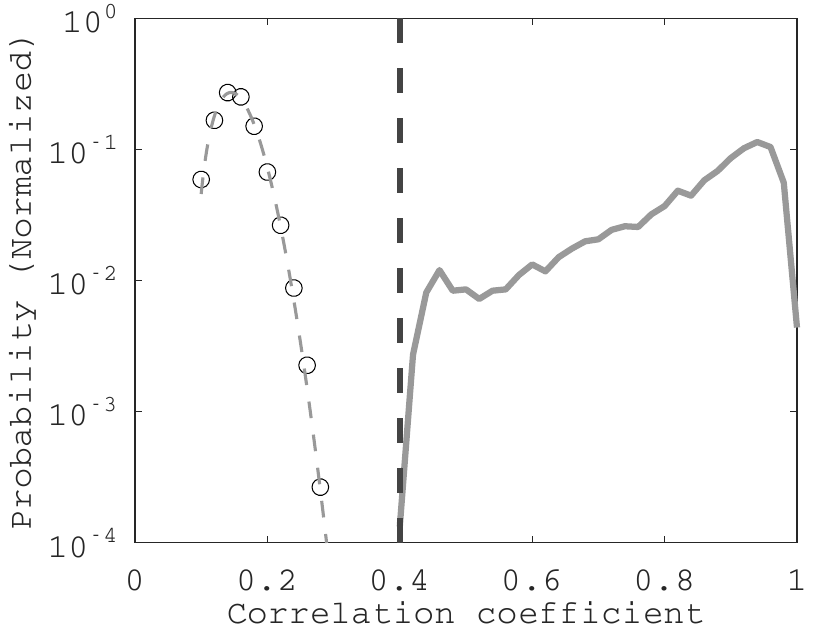}
\caption{\label{fig:fig3} (Black circles) Noise distribution. (Gray dashed line) Fitting function to noise distribution.  (Solid gray line) UHE-$\nu$ signal distribution.  (Dashed black line) Correlation threshold.}
\end{figure}

The Hilbert envelope of the coherently summed waveform (CSW) was calculated from the traces and cross-correlated with Eqs. \ref{eq:final}-\ref{eq:final2}, producing correlation coefficients.  Each coefficient was maximized by varying only $\sigma_t$.  The RF channel properties, $f_0$ and $\gamma$, were held constant.  To find the optimal $f_0$ and $\gamma$, the noise temperature was set to 3 K, so that thermal noise was negligible.  The values for $f_0$ and $\gamma$ that yielded optimal correlation cofficients were 0.15 GHz and 0.025 GHz, respectively. The cross-correlation results are shown in Fig. \ref{fig:fig3}.

\begin{table}
\small
\centering
\begin{tabular}{| c | c | c |}
\hline
\textbf{Parameter} & \textbf{Value} & \textbf{Note} \\
Ice Model & South Pole & 2015 measurements \\
Signal Model & AHRZ2020 & (see \cite{PhysRevD.101.083005}) \\
Trigger & 3 of 8 channels & $\pm 3v_{\rm rms}$ \\
RF channels & 8 & RF bicone (in firn) \\
Channel filters & [80-1000] MHz & Passband \\
Noise Temperature & 233K & Sets $v_{\rm rms}$ \\
Sampling Rate & 1 GHz & $f_{\rm c} = 500$ MHz \\
Samples per channel & 256 & total time, $256$ ns \\
Channel depths & [-4,-6,-8,...-18] m & cable delays included \\
RF cable type & LMR-400 & $\approx -1$ dB at 20 m  \\
\hline
\end{tabular}
\caption{\label{tab:1} Important NuRadioMC parameters.}
\end{table}

\begin{figure}
\centering
\includegraphics[width=0.5\textwidth,trim=3.25cm 8.25cm 4.5cm 9.0cm,clip=true]{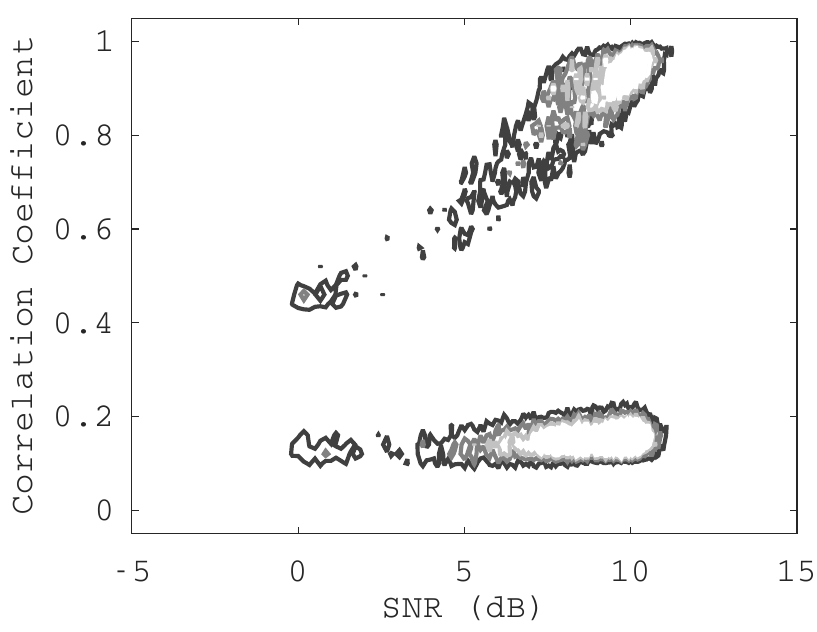}
\caption{\label{fig:fig4} The correlation versus SNR (dB) for UHE-$\nu$ signals (upper distribution) and RF thermal noise (lower distribution).  Color scale: normalized histogram value, with five equally spaced contours between 0.0 and 0.002.}
\end{figure}

\begin{figure}
\centering
\includegraphics[width=0.5\textwidth]{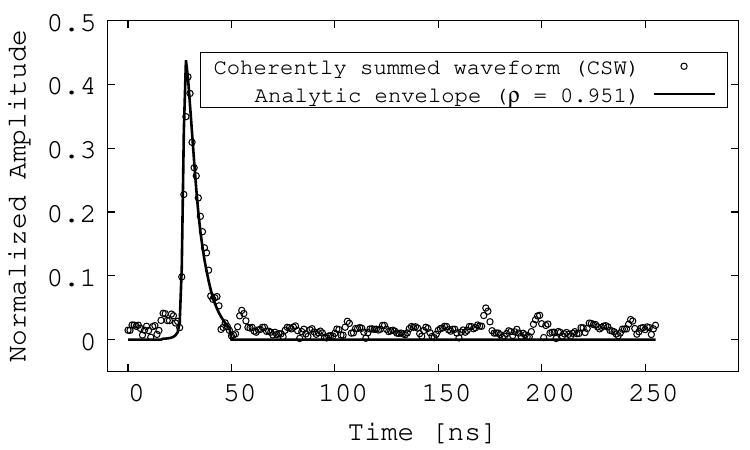}
\caption{\label{fig:example_waveforms} A single-pulse CSW signal (black dots) from a 100 PeV UHE-$\nu$, matched to an analytic envelope (thick black line) with a correlation coefficient $\rho = 0.951$.}
\end{figure}

In Fig. \ref{fig:fig3}, the circles represent the normalized histogram of the correlation coefficient between the optimized analytic envelope and thermal noise.  A fitting function of the form $x^2 \exp(-0.5 x^2)$ was fit to the noise distribution, and is represented by the gray dashed line.  The solid gray line represents the signal distribution, which peaks at a correlation value of 0.94.  Lower signal correlation values correspond to lower SNR values (Fig. \ref{fig:fig4}).  The vertical black dashed line represents a threshold of 0.4.  For the simulated UHE-$\nu$, 99.99\% of correlations exceed this threshold.  Assuming a thermal trigger rate of 1 Hz, integrating the PDF of the noise distribution above the correlation threshold is equivalent to 0.2 noise events every 5 years.  An example CSW fit by the analytic envelope is shown in Fig. \ref{fig:example_waveforms}.

To ensure this result is interpretable, we studied the effect of our NuRadioMC trigger threshold on the thermal trigger rate.  The \textit{in-situ} stations of ARIANNA, ARA, and RNO-G all use majority logic or phased array triggers that record thermal noise events \cite{10.1088/1475-7516/2020/03/053,allison2022lowthreshold-795,agarwal2024instrument-844}.  Individual channels must satisfy the majority logic within a specified gate time.  When a thermal noise fluctuation satisfies the majority logic, the system records data, and a dead-time is required to reset the data acquisition system.

We created a computational tool to estimate the thermal trigger rate.  With a majority logic of 3 of 8, a dead-time of 10 ms, a gate time of 200 ns (the default setting in NuRadioMC), sampling rate of 1 GHz, and thresholds of $\approx \pm 3.5 v_{\rm rms}$, we found that Gaussian white noise triggered at $\approx 1$ Hz.  The difference between thresholds of $\pm 3.5 v_{\rm rms}$ and $\approx \pm 3.0 v_{\rm rms}$ is negligible for fitting Eqs. \ref{eq:final}-\ref{eq:final2} to CSWs from UHE-$\nu$ events.  Though we found that such thermal trigger rates allow just 0.2 thermal noise events to pass the correlation threshold in 5 years, this result can be scaled to the appropriate thermal rate, given the details in \cite{10.1088/1475-7516/2020/03/053,allison2022lowthreshold-795,agarwal2024instrument-844}.

The correlation between the optimized analytical envelope and UHE-$\nu$ signals depends on the SNR (dB).  Let $v_{\rm pp}$ represent the peak-to-peak value in the voltage trace.  The SNR (dB) is

\begin{equation}
\snr_{\rm dB} = 20\log_{10}\left(\frac{1}{2}\frac{v_{\rm pp}}{v_{\rm rms}}\right)
\end{equation}

In Fig. \ref{fig:fig4}, the correlation coefficient is plotted versus the SNR in dB for the data shown in Fig. \ref{fig:fig3}.  The upper and lower distributions correspond to CSWs from UHE-$\nu$ events and thermal noise, respectively.  All RF thermal noise events satisfy the station trigger.  Note that the correlation coefficient for UHE-$\nu$ events is proportional to SNR$_{\rm dB}$.  The SNR of a CSW does not equal the SNR of the individual voltage traces.  Rather, the voltage traces will have SNRs 5-9 dB \textit{lower} than the CSW.  If $N$ voltage traces contain signal, computing the CSW raises the linear SNR by a factor of $\sqrt{N}$, and adds $10\log_{10}(N)$ to SNR$_{\rm dB}$.  For an event with 3 of 8 channels containing signal, $10\log_{10}(3)\approx 5$ dB, while $10\log_{10}(8)\approx 9$ dB.


Our technique has may be applied in a variety of ways.  Though we compared CSWs to Eqs. \ref{eq:final}-\ref{eq:final2} for this analysis,  Eqs. \ref{eq:final}-\ref{eq:final2} can also be matched to individual RF channels.  Future work could include a nuanced analysis that only includes RF channels in the CSW that meet a SNR$_{\rm dB}$ criteria.  Further, our technique applies to all types of RF dipole channels.  (The specific one used in our analysis is in Tab. \ref{tab:1}).  The only requirement for the RF channel is that $r(t)$ from Sec. \ref{sec:onc} represents a good model of the impulse response, and that suitable values for $f_0$ and $\gamma$ can be measured.  An important use case is to model the RF phased-array channels of RNO-G \cite{agarwal2024instrument-844}.  In the seven deployed stations, the RF dipole channels are located at depths that ensure the index of refraction is constant, meaning $r(t)$ should serve at least as well as it did for the RF channels of the RICE detector \cite{10.1103/PhysRevD.85.062004}.

\section{Measurement of $\log_{10}E_{\rm C}$}
\label{sec:recon}

In the following analysis, we summarize a technique that yields a measurement of the logarithm of the UHE-$\nu$ cascade energy, $\ln\Lambda$, using Eqs. \ref{eq:uncert}-\ref{eq:a_err}.  For Eq. \ref{eq:uncert}, $\sigma_t$ is measured from the optimized analytic envelope, $c$ and $\theta_{\rm C}$ are known constants.  The $\Delta\theta$ must be measured separately from $\sigma_t$, or assumption must be made about its value.  Suppose that $\Delta\theta$ is distributed normally, with zero mean, so that $\Delta\theta \approx \Delta\theta_{\rm rms}$.  Solving Eq. \ref{eq:uncert} for $a$ gives

\begin{equation}
a = \frac{c\sigma_t}{\Delta\theta_{\rm rms} \sin\theta_{\rm C}}
\end{equation}

The result for the fractional error in $a$ is found by propagating error from $\sigma_t$ and $\Delta\theta$, defined as $\epsilon$ and $\sigma_{\Delta\theta}$, respectively.  The result is

\begin{equation}
\frac{\sigma_a}{a} = \left(\left(\frac{\epsilon}{\sigma_t}\right)^2 +  \left(\frac{\sigma_{\Delta\theta}}{\Delta\theta}\right)^2\right)^{1/2} \label{eq:err_prop}
\end{equation}

The first term is small compared to the second, as it is limited by the scan resolution for $\sigma_t$ and the number of samples per analytic envelope.  In Sec. \ref{sec:sim}, the scan resolution was set to 0.2 ns.  There are typically $>10$ samples per envelope, or about 10 ns.  The fractional error in $\Delta\theta$ would be $1.0$, because the rms and $\sigma$ values are equal for normal distributions with zero mean.  Thus, $\left(\epsilon/\sigma_t\right)^2$ is two orders of magnitude smaller than $\left(\sigma_{\Delta\theta}/\Delta\theta\right)^2$, so

\begin{equation}
\frac{\sigma_a}{a} = \left|\frac{\sigma_{\Delta\theta}}{\Delta\theta}\right| = \left|\frac{\Delta\theta_{\rm rms}}{\Delta\theta_{\rm rms}}\right|\approx 1
\end{equation}

Inserting $(\sigma_a/a)\approx 1$ into Eq. \ref{eq:a_err} gives

\begin{equation}
\frac{\sigma_{\ln\Lambda}}{\ln\Lambda} \approx 2 \label{eq:a_err_2}
\end{equation}

Using Eqs. \ref{eq:em} and \ref{eq:had}, the logarithm of the energy is

\begin{equation}
\ln\Lambda = \left( \frac{c\sigma_t}{x_{\rm em/had} \Delta\theta_{\rm rms}\sin\theta_{\rm C}} \right)^2 \label{eq:lnLambda}
\end{equation}

Using Eqs. 10 and 12 from HH2022 (\cite{PhysRevD.105.123019}), $x_{\rm em}$ and $x_{\rm had}$ were found to be 0.80 and 0.93 meters, respectively (FWHM, $R=0.5$).  Using Eq. \ref{eq:lnLambda}, the $\sigma_t$ results from the optimized envelope fits to UHE-$\nu$ signals could be used to deduce the logarithm of the UHE-$\nu$ cascade, $\log_{10} E_{\rm C}$.  First, converting to the base-10 logarithm introduces a factor of $\ln(10)$ in the denominator of Eq. \ref{eq:lnLambda}.  Second, $\ln\Lambda = \ln(E_{\rm C}/E_{\rm crit})$, where $E_{\rm C}$ is the cascade energy, and $E_{\rm crit}\approx 10^8$ eV is known as the critical energy \cite{PhysRevD.105.123019}.  Since $\ln\Lambda = \ln E_{\rm C} - \ln E_{\rm crit}$, separating this ratio adds $\log_{10} E_{\rm crit}$ to the right hand side of Eq. \ref{eq:lnLambda}.  Third, let $x_{\rm ave}$ be the average of $x_{\rm em}$ and $x_{\rm had}$, to reflect the unknown cascade type.  The modified form of Eq. \ref{eq:lnLambda} is

\begin{equation}
\log_{10} E_{\rm C} = \frac{\left( c\sigma_t \right)^2}{\ln 10(x_{\rm ave} \Delta\theta_{\rm rms}\sin\theta_{\rm C})^2} + \log_{10} E_{\rm crit} \label{eq:lnLambda_2}
\end{equation}

In hopes of using Eq. \ref{eq:lnLambda_2} to measure $\log_{10} E_{\rm C}$, we calculated the $\sigma_t$ distribution derived from 15133 UHE-$\nu$ events that triggered the detector with $E_{\rm C} = 100$ PeV.  Though the $\sigma_t$ distribution corresponds to the correlation coefficient distribution in Fig. \ref{fig:fig3}, two effects conspire to cause the measurement of $\log_{10} E_{\rm C}$ to break down.  First, the dependence of $\log_{10} E_{\rm C}$ in Eq. \ref{eq:lnLambda_2} on $\sigma_t$ is \textit{quadratic}, so subtle shifts in $\sigma_t$ amplify error in $\log_{10} E_{\rm C}$.  Second, \textit{reflected signals} lead to extended tails in the $\sigma_t$ distribution that skew measurements of the average of $\log_{10} E_{\rm C}$.

Reflected signals occur within traces either when signals from two ray-tracing paths are available, or when part of the signal reflects from the surface back to the RF channel.  Signals that propagate directly to the detector are known as \textit{direct signals}, while signals that reflect at the snow-air interface are known as \textit{reflected signals}.  Complex RF signal propagation in polar ice has been studied, and even proposed as a UHE-$\nu$ energy reconstruction technique \cite{Barwick:2018497,anker2019neutrino-734,deaconu2018measurements-182}.

\begin{figure}
\centering
\includegraphics[width=0.5\textwidth]{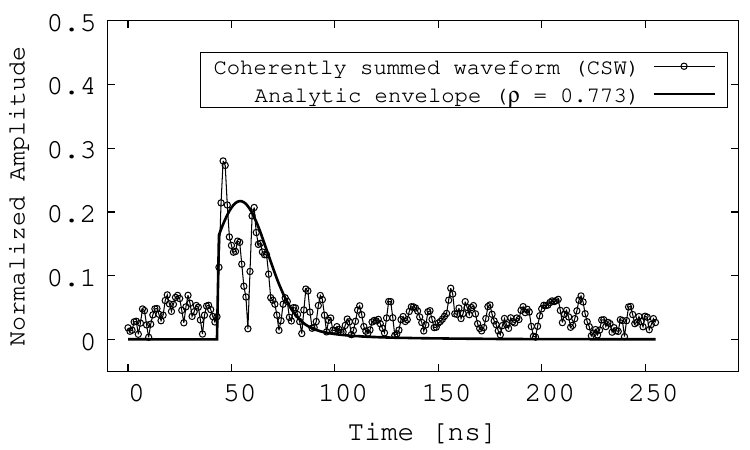}
\caption{\label{fig:example_waveforms_2} A CSW signal with a reflection (black dots and thin black line) from a 100 PeV UHE-$\nu$, matched to an analytic envelope (thick black line), with $\rho = 0.773$.}
\end{figure}

When reflected signals are present in the CSW, the fitting algorithm compensates with unphysical $\sigma_t$ values corresponding to a single pulse (see Fig. \ref{fig:example_waveforms_2}).  This compensation leads to an overestimation of $\log_{10} E_{\rm C}$.  To account for reflected signals, we would have to assume two independent copies of Eqs. \ref{eq:final}-\ref{eq:final3} for each UHE-$\nu$ event, and scan for two $\sigma_t$ values and two pulse locations within the waveform.  Though not conceptually difficult, we felt the technique was beyond the scope of this work.


\section{Conclusion}
\label{sec:conc}

In Secs. \ref{sec:int} and \ref{sec:unit}, previous work in the utility of analytic Askaryan models was reviewed, and important parameters were defined.  In Sec. \ref{sec:onc}, a fully analytic model for $\mathcal{E}_{\rm s*r}(t)$ and $s(t) * r(t)$ was presented and checked for accuracy.  In Sec. \ref{sec:sim}, the analytic model was shown to reject all but 0.2 RF thermal triggers in a 5 year run at a 1 Hz thermal trigger rate using a correlation threshold of $\rho\geq 0.4$.  The threshold $\rho\geq 0.4$ preserves 99.99\% of UHE-$\nu$ signals at 100 PeV.  The correlation distribution for UHE-$\nu$ signals peaked at 0.94, and some values even exceed 0.94.  Finally, in Sec. \ref{sec:recon}, we sketched a technique to calculate the average $\log_{10} E_{\rm C}$ from data from \textit{a single-string detector}.  The technique breaks down without explicitly accounting for signal reflections.

There are three logical steps for this work in the future.  First, we must account for reflected signals  when fitting analytic envelopes to CSWs from UHE-$\nu$.  The authors of \cite{anker2019neutrino-734} have shown how direct and reflected (DnR) signals may be used to constrain the UHE-$\nu$ energy.  This procedure should be combined with the analytic envelope fit to advance UHE-$\nu$ energy reconstruction.  Though this is not difficult, it was beyond the scope of this work.

Second, a separate fit for $\Delta\theta$ should be included.  This parameter may be constrained by using the spectral cutoff of the voltage traces, which are inversely correlated to $\Delta\theta$ \cite{10.1016/j.astropartphys.2017.03.008}.  With a separate fit for $\Delta\theta$, the $a$-value could be constrained without making any assumption about $\Delta\theta$.  This step would improve the precision of the $a$-value by lowering the fractional error in $a$-values in Eq. \ref{eq:err_prop}.  This procedure was completed in \cite{PhysRevD.105.123019}, but it is restricted to comparisons between $\vec{E}$-fields and not observed CSWs.

Third, it is interesting to note that the model in this work relies on the \textit{off-cone} ($\theta \neq \theta_{\rm C}$) version of $s(t)$.  The authors of HH2022 \cite{PhysRevD.105.123019} also include an \textit{on-cone} version of $s(t)$ that is used when $\Delta\theta = 0$, and can be used as $\Delta\theta\to 0$.  The on-cone model could provide a better fit between $\mathcal{E}_{\rm s*r}(t)$ for events with $\Delta\theta \lesssim 1$ degree \cite{PhysRevD.105.123019}.  Further, the on-cone model is defined in a piecewise fashion by rising and falling exponential functions, simplifying the procedure used to arrive at Eqs. \ref{eq:final}-\ref{eq:final2}.

We would like to thank Prof. Steve Barwick, and Prof. Mark Kozek for useful background discussions.  We are grateful to the authors of RB, Profs. John Ralston and Roman Buniy, for inspiring us in this line of research \cite{10.1103/physrevd.65.016003}.  We would like to thank Prof. Dave Besson for encouragement and wisdom while we continued to develop the model.  Most of all, we would like to thank Prof. Amy Connolly for inspiring us and encouraging us to master more advanced applications of complex analysis to mathematical physics.  We hope the model will help identify the first UHE-$\nu$ via the Askaryan effect, especially in light of potential UHE-$\nu$ observations \cite{collaboration2025observation-22f}.

\appendix

\section{Python3 Code for $\mathcal{L}\lbrace D(u-x)\rbrace_k$}
\label{sec:appA}

The code in this section computes the Laplace transform of the shifted Dawson function, $\mathcal{L}\lbrace D(u-x)\rbrace_k$.  The syntax has been adjusted to fit in two-column format, so please check the alignment.

\small
\begin{verbatim}
import numpy as np
from scipy.integrate import quad
from scipy.special import dawsn as D
...
def shifted_laplace_transform(self,f,s,t0,t_max=1000):
    x = s.real
    y = s.imag
    re_int = quad(lambda t: 
        np.exp(-x*t)*np.cos(y*t)*f(t-t0),0,t_max)[0]
    im_int = quad(lambda t: 
        np.exp(-x*t)*np.sin(y*t)*f(t-t0),0,t_max)[0]
    return complex(re_int,-im_int)
...
x = t/(sigma_t*np.sqrt(2))
z = (2*np.pi*J*f0-2*np.pi*gamma)*np.sqrt(2)*sigma_t
k=-z
... + k*self.shifted_laplace_transform(D,k,x))
\end{verbatim}

\section{Python3 Example Code for $\mathcal{E}_{\rm s * r}(t)$}
\label{sec:appB}

The code in this section computes $\mathcal{E}_{\rm s * r}(t)$.  The syntax has been adjusted to fit in two-column format, so please check the alignment.

\small
\begin{verbatim}
import numpy as np
from scipy.integrate import quad
from scipy.special import wofz as w
from scipy.special import dawsn as D
...
def shifted_laplace_transform(...):
...
def dwdq(q):
    J = complex(0,1)
    return 2*J/np.sqrt(np.pi)-2*q*w(q)
def math_env(self,t,E0,R0,sigma_t,f0,gamma):
    units = R0*E0*sigma_t*sigma_t
    J = complex(0,1)
    x = t/(sigma_t*np.sqrt(2))
    z = (2*np.pi*J*f0-2*np.pi*gamma)
    z *= np.sqrt(2)*sigma_t
    k=-z
    q = -J*(x+z/2)
    first = -np.sqrt(np.pi)*(x*np.exp(-x*x)*w(q)
        +0.5*J*np.exp(-x*x)*self.dwdq(q))
    second = 2/np.sqrt(np.pi)*(D(x)
        +k*self.shifted_laplace_transform(D,k,x))
    if(np.isinf(first) or np.isinf(second) 
        or np.isnan(first) or np.isnan(second)):
        return 0.0
    else:
        result = units*(first+J*second)
    return 0.5*np.abs(result)
\end{verbatim}

\section{Python3 Example Code for $s(t) * r(t)$}
\label{sec:appC}

The code in this section contains the definitions of $s(t)$ and $r(t)$, and computes $s(t) * r(t)$.  The syntax has been adjusted to fit in two-column format, so please check the alignment.

\small
\begin{verbatim}
import numpy as np
from scipy.special import wofz as w
...
def dwdq(...):
...
def s(t,E0,sigma_t):
    return -E0*t*np.exp(-0.5*t*t/sigma_t/sigma_t)
def r(t,R0,f0,gamma):
    if(t>=0):
        result = R0*np.cos(2*np.pi*f0*t)
        result *= np.exp(-2*np.pi*gamma*t)
        return result
    else:
        return 0
def math_conv(t,E0,R0,sigma_t,f0,gamma):
    J = complex(0,1)
    units = R0*E0*sigma_t*sigma_t
    x = t/(sigma_t*np.sqrt(2))
    z = (2*np.pi*J*f0-2*np.pi*gamma)
    z *= np.sqrt(2)*sigma_t
    q = -J*(x+0.5*z)
    Re_wq = np.real(w(q))
    Re_dwdx = np.real(-J*dwdq(q))
    result = x*np.exp(-x*x)*Re_wq
    result -= 0.5*np.exp(-x*x)*Re_dwdx)
    if(np.isinf(result) or np.isnan(result)):
        return 0.0
    else:
        return -np.sqrt(np.pi)*units*result
\end{verbatim}

\bibliography{apssamp}

@article{allison2022lowthreshold-795, 
  year     = {2022}, 
  title    = {Low-threshold ultrahigh-energy neutrino search with the Askaryan Radio Array}, 
  author   = {Allison, P and Archambault, S and Beatty, J J and Besson, D Z and Bishop, A and Chen, C C and Chen, C H and Chen, P and Chen, Y C and Clark, B A and Clay, W and Connolly, A and Cremonesi, L and Dasgupta, P and Davies, J and Kockere, S de and Vries, K D de and Deaconu, C and {DuVernois}, M A and Flaherty, J and Friedman, E and Gaior, R and Hanson, J and Harty, N and Hendricks, B and Hoffman, K D and Hong, E and Hsu, S Y and Hu, L and Huang, J J and Huang, M -H and Hughes, K and Ishihara, A and Karle, A and Kelley, J L and Kim, K -C and Kim, M -C and Kravchenko, I and Krebs, R and Ku, Y and Kuo, C Y and Kurusu, K and Landsman, H and Latif, U A and Li, C -J and Liu, T -C and Lu, M -Y and Madison, B and Madison, K and Mase, K and Meures, T and Nam, J and Nichol, R J and Nir, G and Novikov, A and Nozdrina, A and Oberla, E and Osborn, J and Pan, Y and Pfendner, C and Punsuebsay, N and Roth, J and Seckel, D and Seikh, M F H and Shiao, Y -S and Shultz, A and Smith, D and Toscano, S and Torres, J and Touart, J and Eijndhoven, N van and Varner, G S and Vieregg, A and Wang, M -Z and Wang, S -H and Wang, Y H and Wissel, S A and Xie, C and Yoshida, S and Young, R and Collaboration, {ARA}}, 
  journal  = {Physical Review D}, 
  issn     = {2470-0010}, 
  doi      = {10.1103/physrevd.105.122006}, 
  eprint   = {2202.07080}, 
  pages    = {122006}, 
  number   = {12}, 
  volume   = {105}
}

@article{agarwal2024instrument-844, 
  year     = {2024}, 
  title    = {Instrument design and performance of the first seven stations of {RNO}-G}, 
  author   = {Agarwal, S and Aguilar, J A and Alden, N and Ali, S and Allison, P and Betts, M and Besson, D and Bishop, A and Botner, O and Bouma, S and Buitink, S and Camphyn, R and Cataldo, M and Chiche, S and Clark, B A and Coleman, A and Couberly, K and Kockere, S de and Vries, K D de and Deaconu, C and Glaser, C and Glüsenkamp, T and Hallgren, A and Hallmann, S and Hanson, J C and Hendricks, B and Henrichs, J and Heyer, N and Hornhuber, C and Hughes, K and Karg, T and Karle, A and Kelley, J L and Kerr, C and Klein, C and Korntheuer, M and Kowalski, M and Kravchenko, I and Krebs, R and Lahmann, R and Latif, U and Laub, P and Liu, C -H and Marsee, M J and Meyers, Z S and Mikhailova, M and Mulrey, K and Muzio, M and Nelles, A and Novikov, A and Nozdrina, A and Oberla, E and Oeyen, B and Polfrey, S and Punsuebsay, N and Pyras, L and Ravn, M and Reichert, M and Rix, J and Ryckbosch, D and Schlüter, F and Scholten, O and Seckel, D and Seikh, M F H and Smith, D and Stoffels, J and Terveer, K and Toscano, S and Tosi, D and Tutt, J and Broeck, D J Van Den and Eijndhoven, N van and Vieregg, A G and Vijai, A and Welling, C and Williams, D R and Windischhofer, P and Veale, J and Wissel, S and Young, R and Zink, A}, 
  journal  = {{arXiv}}, 
  doi      = {10.48550/arxiv.2411.12922}, 
  eprint   = {2411.12922}
}

@article{Barwick_Besson_Gorham_Saltzberg_2005,
	title={South Polar in situ radio-frequency ice attenuation},
	volume={51},
	DOI={10.3189/172756505781829467},
	number={173},
	journal={Journal of Glaciology},
	author={Barwick, S. and Besson, D. and Gorham, P. and Saltzberg, D.},
	year={2005}, 
	pages={231–238}
}

@article{welling2024brief-b47, 
  year     = {2024}, 
  title    = {Brief communication: Precision measurement of the index of refraction of deep glacial ice at radio frequencies at Summit Station, Greenland}, 
  author   = {Welling, Christoph and Collaboration, The {RNO}-G}, 
  journal  = {The Cryosphere}, 
  doi      = {10.5194/tc-18-3433-2024}, 
  pages    = {3433--3437}, 
  number   = {7}, 
  volume   = {18}
}

@article{deaconu2018measurements-182, 
  year     = {2018}, 
  title    = {Measurements and modeling of near-surface radio propagation in glacial ice and implications for neutrino experiments}, 
  author   = {Deaconu, C. and Vieregg, A. G. and Wissel, S. A. and Bowen, J. and Chipman, S. and Gupta, A. and Miki, C. and Nichol, R. J. and Saltzberg, D.}, 
  journal  = {Physical Review D}, 
  issn     = {2470-0010}, 
  doi      = {10.1103/physrevd.98.043010}, 
  eprint   = {1805.12576}, 
  pages    = {043010}, 
  number   = {4}, 
  volume   = {98}
}

@article{anker2019neutrino-734, 
  year     = {2019}, 
  title    = {Neutrino vertex reconstruction with in-ice radio detectors using surface reflections and implications for the neutrino energy resolution}, 
  author   = {Anker, A. and Barwick, S.W. and Bernhoff, H. and Besson, D.Z. and Bingefors, N. and García-Fernández, D. and Gaswint, G. and Glaser, C. and Hallgren, A. and Hanson, J.C. and Klein, S.R. and Kleinfelder, S.A. and Lahmann, R. and Latif, U. and Nam, J. and Novikov, A. and Nelles, A. and Paul, M.P. and Persichilli, C. and Plaisier, I. and Prakash, T. and Shively, S.R. and Tatar, J. and Unger, E. and Wang, S.-H. and Welling, C. and Zierke, S.}, 
  journal  = {Journal of Cosmology and Astroparticle Physics}, 
  doi      = {10.1088/1475-7516/2019/11/030}, 
  eprint   = {1909.02677}, 
  pages    = {030--030}, 
  number   = {11}, 
  volume   = {2019}
}

@article{collaboration2025observation-22f, 
  year     = {2025}, 
  title    = {Observation of an ultra-high-energy cosmic neutrino with {KM}3NeT.}, 
  author   = {{The {KM}3NeT Collaboration}}, 
  journal  = {Nature}, 
  issn     = {0028-0836}, 
  doi      = {10.1038/s41586-024-08543-1}, 
  pmid     = {39939793}, 
  pmcid    = {{PMC}11821517}, 
  pages    = {376--382}, 
  number   = {8050}, 
  volume   = {638}
}

@article{collaboration2023observation-08b, 
  year     = {2023}, 
  title    = {Observation of high-energy neutrinos from the Galactic plane}, 
  author   = {{The IceCube Collaboration}}, 
  journal  = {Science}, 
  issn     = {0036-8075}, 
  doi      = {10.1126/science.adc9818}, 
  pmid     = {37384687}, 
  pages    = {1338--1343}, 
  number   = {6652}, 
  volume   = {380}
}

@article{collaboration2022evidence-a08, 
  year     = {2022}, 
  title    = {Evidence for neutrino emission from the nearby active galaxy {NGC} 1068}, 
  author   = {{The IceCube Collaboration}}, 
  journal  = {Science}, 
  issn     = {0036-8075}, 
  doi      = {10.1126/science.abg3395}, 
  pmid     = {36378962}, 
  pages    = {538--543}, 
  number   = {6619}, 
  volume   = {378}
}

@article{collaboration2021detection-6fa, 
  year     = {2021}, 
  title    = {Detection of a particle shower at the Glashow resonance with {IceCube}}, 
  author   = {{The IceCube Collaboration}}, 
  journal  = {Nature}, 
  issn     = {0028-0836}, 
  doi      = {10.1038/s41586-021-03256-1}, 
  pmid     = {33692563}, 
  eprint   = {2110.15051}, 
  pages    = {220--224}, 
  number   = {7849}, 
  volume   = {591}
}

@article{collaboration2018neutrino-2a0, 
  year     = {2018}, 
  title    = {Neutrino emission from the direction of the blazar {TXS} 0506+056 prior to the {IceCube}-170922A alert}, 
  author   = {{The IceCube Collaboration}}, 
  journal  = {Science}, 
  issn     = {0036-8075}, 
  doi      = {10.1126/science.aat2890}, 
  pmid     = {30002248}, 
  eprint   = {1807.08794}, 
  pages    = {147--151}, 
  number   = {6398}, 
  volume   = {361}
}

@article{collaboration2016observation-03b, 
  year     = {2016}, 
  title    = {{OBSERVATION} {AND} {CHARACTERIZATION} {OF} A {COSMIC} {MUON} {NEUTRINO} {FLUX} {FROM} {THE} {NORTHERN} {HEMISPHERE} {USING} {SIX} {YEARS} {OF} {ICECUBE} {DATA}}, 
  author   = {{The IceCube Collaboration}}, 
  journal  = {The Astrophysical Journal}, 
  issn     = {0004-637X}, 
  doi      = {10.3847/0004-637x/833/1/3}, 
  eprint   = {1607.08006}, 
  pages    = {3}, 
  number   = {1}, 
  volume   = {833}
}

@article{aartsen2013first-bff, 
  year     = {2013}, 
  title    = {First Observation of {PeV}-Energy Neutrinos with {IceCube}}, 
  author   = {{The IceCube Collaboration}}, 
  journal  = {Physical Review Letters}, 
  issn     = {0031-9007}, 
  doi      = {10.1103/physrevlett.111.021103}, 
  pmid     = {23889381}, 
  eprint   = {1304.5356}, 
  pages    = {021103}, 
  number   = {2}, 
  volume   = {111}
}

@article{10.48550/arxiv.2008.04323, 
year = {2020}, 
title = {{IceCube-Gen2: The Window to the Extreme Universe}}, 
author = {{The IceCube-Gen2 Collaboration}}, 
journal = {arXiv}, 
doi = {10.48550/arxiv.2008.04323}, 
eprint = {2008.04323}
}

@article{10.1088/1748-0221/16/03/p03025, 
year = {2021}, 
title = {{Design and sensitivity of the Radio Neutrino Observatory in Greenland (RNO-G)}}, 
author = {{The ARA Collaboration}},
journal = {Journal of Instrumentation}, 
doi = {10.1088/1748-0221/16/03/p03025}, 
eprint = {2010.12279}, 
pages = {P03025}, 
number = {03}, 
volume = {16}
}

@article{10.1063/1.4822832, 
year = {1989}, 
title = {{Dawson's Integral and the Sampling Theorem}}, 
author = {Rybicki, George B}, 
journal = {Computers in Physics}, 
issn = {0894-1866}, 
doi = {10.1063/1.4822832}, 
pages = {85--87}, 
number = {2}, 
volume = {3}
}

@article{PhysRevD.105.123019,
  title = {Complex analysis of Askaryan radiation: A fully analytic model in the time domain},
  author = {Hanson, Jordan C. and Hartig, Raymond},
  journal = {Phys. Rev. D},
  volume = {105},
  issue = {12},
  pages = {123019},
  numpages = {18},
  year = {2022},
  month = {Jun},
  publisher = {American Physical Society},
  doi = {10.1103/PhysRevD.105.123019},
  url = {https://link.aps.org/doi/10.1103/PhysRevD.105.123019}
}

@article{10.1126/science.1242856, 
year = {2013}, 
title = {{Evidence for High-Energy Extraterrestrial Neutrinos at the IceCube Detector}}, 
author = {{The IceCube Collaboration}}, 
journal = {Science}, 
issn = {0036-8075}, 
doi = {10.1126/science.1242856}, 
pmid = {24264993}, 
eprint = {1311.5238}, 
pages = {1242856--1242856}, 
number = {6161}, 
volume = {342}
}

@article{10.1103/physrevd.98.062003, 
year = {2018}, 
title = {{Differential limit on the extremely-high-energy cosmic neutrino flux in the presence of astrophysical background from nine years of IceCube data}}, 
author = {{The IceCube Collaboration}},
journal = {Physical Review D}, 
issn = {2470-0010}, 
doi = {10.1103/physrevd.98.062003}, 
eprint = {1807.01820}, 
pages = {062003}, 
number = {6}, 
volume = {98}
}

@article{10.1088/1475-7516/2020/03/053, 
year = {2020}, 
title = {{A search for cosmogenic neutrinos with the ARIANNA test bed using 4.5 years of data}}, 
author = {{The ARIANNA Collaboration}}, 
journal = {Journal of Cosmology and Astroparticle Physics}, 
doi = {10.1088/1475-7516/2020/03/053}, 
eprint = {1909.00840}, 
pages = {053--053}, 
number = {03}, 
volume = {2020}
}

@article{10.1103/physrevd.102.043021, 
year = {2020}, 
title = {{Constraints on the diffuse flux of ultrahigh energy neutrinos from four years of Askaryan Radio Array data in two stations}}, 
author = {{The ARA Collaboration}}, 
journal = {Physical Review D}, 
issn = {2470-0010}, 
doi = {10.1103/physrevd.102.043021}, 
eprint = {1912.00987}, 
pages = {043021}, 
number = {4}, 
volume = {102}
}

@article{10.3189/2015jog14j214, 
year = {2015}, 
title = {{Radar absorption, basal reflection, thickness and polarization measurements from the Ross Ice Shelf, Antarctica}}, 
author = {{J. C. Hanson et al}}, 
journal = {Journal of Glaciology}, 
issn = {0022-1430}, 
doi = {10.3189/2015jog14j214}, 
pages = {438--446}, 
number = {227}, 
volume = {61}
}

@article{10.3189/2015jog15j057, 
year = {2014},
journal = {Journal of Glaciology}, 
title = {{An in situ measurement of the radio-frequency attenuation in ice at Summit Station, Greenland}}, 
author = {Avva, J and Kovac, JM and Miki, C and Saltzberg, D and Vieregg, AG}, 
doi = {10.3189/2015jog15j057}, 
eprint = {1409.5413}
}

@article{allison2012design-cd3, 
  year     = {2012}, 
  title    = {Design and initial performance of the Askaryan Radio Array prototype {EeV} neutrino detector at the South Pole}, 
  author   = {Allison, Patrick and Auffenberg, Jan and Bard, Robert and Beatty, {JJ} and Besson, {DZ} and Böser, S and Chen, C and Chen, Pisin and Connolly, Amy and Davies, Jonathan}, 
  journal  = {Astroparticle Physics}, 
  issn     = {0927-6505}, 
  doi      = {10.1016/j.astropartphys.2011.11.010}, 
  eprint   = {1105.2854}, 
  pages    = {457--477}, 
  number   = {7}, 
  volume   = {35}
}

@article{zhs, 
year = {1992}, 
title = {{Electromagnetic pulses from high-energy showers: Implications for neutrino detection}}, 
author = {Zas, E and Halzen, Francis and Stanev, T}, 
journal = {Physical Review D}, 
pmid = {10014221}, 
pages = {362}, 
number = {1}, 
volume = {45}
}

@article{10.1103/PhysRevD.85.062004, 
year = {2012}, 
title = {{Updated results from the RICE experiment and future prospects for ultra-high energy neutrino detection at the south pole}}, 
author = {{I. Kravchenko et al}}, 
journal = {Physical Review D}, 
issn = {2470-0029}, 
doi = {10.1103/PhysRevD.85.062004}, 
eprint = {1106.1164}, 
pages = {062004}, 
number = {6}, 
volume = {85}
}

@article{saltzberg, 
year = {2001}, 
title = {{Observation of the Askaryan effect: coherent microwave Cherenkov emission from charge asymmetry in high-energy particle cascades.}}, 
author = {Saltzberg, D and Gorham, P and Walz, D and Field, C and Iverson, R and Odian, A and Resch, G and Schoessow, P and Williams, D}, 
journal = {Physical review letters}, 
issn = {0031-9007}, 
pmid = {11290043}, 
pages = {2802--5}, 
number = {13}, 
volume = {86}
}

@article{10.1103/PhysRevD.74.043002, 
year = {2006}, 
title = {{Time-domain measurement of broadband coherent Cherenkov radiation}}, 
author = {Miocinovic, P and Field, RC and Gorham, PW and Guillian, E and Milincic, R and Saltzberg, D and Walz, D and Williams, D}, 
journal = {Physical Review D}, 
issn = {2470-0029}, 
doi = {10.1103/PhysRevD.74.043002}, 
eprint = {hep-ex/0602043}, 
pages = {043002}, 
number = {4}, 
volume = {74}
}

@article{ask_ice,
  title = {Observations of the Askaryan Effect in Ice},
  author = {Gorham, P. W. and Barwick, S. W. and Beatty, J. J. and Besson, D. Z. and Binns, W. R. and Chen, C. and Chen, P. and Clem, J. M. and Connolly, A. and Dowkontt, P. F. and DuVernois, M. A. and Field, R. C. and Goldstein, D. and Goodhue, A. and Hast, C. and Hebert, C. L. and Hoover, S. and Israel, M. H. and Kowalski, J. and Learned, J. G. and Liewer, K. M. and Link, J. T. and Lusczek, E. and Matsuno, S. and Mercurio, B. and Miki, C. and Mio\ifmmode \check{c}\else \v{c}\fi{}inovi\ifmmode \acute{c}\else \'{c}\fi{}, P. and Nam, J. and Naudet, C. J. and Ng, J. and Nichol, R. and Palladino, K. and Reil, K. and Romero-Wolf, A. and Rosen, M. and Ruckman, L. and Saltzberg, D. and Seckel, D. and Varner, G. S. and Walz, D. and Wu, F.},
  collaboration = {ANITA Collaboration},
  journal = {Phys. Rev. Lett.},
  volume = {99},
  issue = {17},
  pages = {171101},
  numpages = {5},
  year = {2007},
  month = {Oct},
  publisher = {American Physical Society},
  doi = {10.1103/PhysRevLett.99.171101},
  url = {https://link.aps.org/doi/10.1103/PhysRevLett.99.171101}
}

@article{10.1016/j.astropartphys.2017.03.008, 
year = {2017}, 
title = {{Complex analysis of Askaryan radiation: A fully analytic treatment including the LPM effect and Cascade Form Factor}}, 
author = {Hanson, Jordan C. and Connolly, Amy L.}, 
journal = {Astroparticle Physics}, 
issn = {0927-6505}, 
doi = {10.1016/j.astropartphys.2017.03.008}, 
pages = {75--89}, 
volume = {91}
}

@article{10.1103/physrevd.84.103003, 
year = {2011}, 
title = {{Practical and accurate calculations of Askaryan radiation}}, 
author = {Alvarez-Muniz, Jaime and Romero-Wolf, Andrés and Zas, Enrique}, 
journal = {Physical Review D}, 
issn = {2470-0029}, 
doi = {10.1103/physrevd.84.103003}, 
eprint = {1106.6283}, 
pages = {103003}, 
number = {10}, 
volume = {84}
}

@article{10.1140/epjc/s10052-020-7612-8, 
year = {2020}, 
title = {{NuRadioMC: simulating the radio emission of neutrinos from interaction to detector}}, 
author = {{C. Glaser et al}}, 
journal = {The European Physical Journal C}, 
issn = {1434-6044}, 
doi = {10.1140/epjc/s10052-020-7612-8}, 
eprint = {1906.01670}, 
pages = {77}, 
number = {2}, 
volume = {80}
}

@article{Barwick:2018497, 
year = {2018}, 
title = {{Observation of classically `forbidden' electromagnetic wave propagation and implications for neutrino detection.}}, 
author = {{The ARIANNA Collaboration}}, 
journal = {Journal of Cosmology and Astroparticle Physics}, 
doi = {10.1088/1475-7516/2018/07/055},
number = {07}, 
volume = {2018}
}

@article{ALLISON201963,
title = {Measurement of the real dielectric permittivity \epsilon_r of glacial ice},
journal = {Astroparticle Physics},
volume = {108},
pages = {63-73},
year = {2019},
issn = {0927-6505},
doi = {https://doi.org/10.1016/j.astropartphys.2019.01.004},
url = {https://www.sciencedirect.com/science/article/pii/S0927650518301154},
author = {{The ARA Collaboration}}
}

@article{PhysRevD.101.083005,
  title = {Askaryan radiation from neutrino-induced showers in ice},
  author = {Alvarez-Muniz, Jaime and Hansen, P. M. and Romero-Wolf, Andr\'es and Zas, Enrique},
  journal = {Phys. Rev. D},
  volume = {101},
  issue = {8},
  pages = {083005},
  numpages = {16},
  year = {2020},
  month = {Apr},
  publisher = {American Physical Society},
  doi = {10.1103/PhysRevD.101.083005},
  url = {https://link.aps.org/doi/10.1103/PhysRevD.101.083005}
}

@article{10.1103/physrevd.65.016003, 
year = {2001}, 
title = {{Radio detection of high energy particles: Coherence versus multiple scales}}, 
author = {Buniy, Roman V. and Ralston, John P.}, 
journal = {Physical Review D}, 
issn = {2470-0029}, 
doi = {10.1103/physrevd.65.016003}, 
number = {1}, 
volume = {65}
}

@article{10.1088/1748-0221/15/09/p09039, 
year = {2020}, 
title = {{Probing the angular and polarization reconstruction of the ARIANNA detector at the South Pole}}, 
author = {{The ARIANNA Collaboration}}, 
journal = {Journal of Instrumentation}, 
doi = {10.1088/1748-0221/15/09/p09039}, 
eprint = {2006.03027}, 
pages = {P09039--P09039}, 
number = {09}, 
volume = {15}
}

@article{10.1016/j.astropartphys.2014.09.002, 
year = {2015}, 
title = {{Time-domain response of the ARIANNA detector}}, 
author = {{J. C. Hanson et al}}, 
journal = {Astroparticle Physics}, 
issn = {0927-6505}, 
doi = {10.1016/j.astropartphys.2014.09.002}, 
eprint = {1406.0820}, 
pages = {139--151}, 
volume = {62}
}

@misc{NIST:DLMF,
         key = "{\relax DLMF}",
       title = "{\it NIST Digital Library of Mathematical Functions}",
howpublished = "http://dlmf.nist.gov/, Release 1.1.1 of 2021-03-15",
         url = "http://dlmf.nist.gov/",
        note = "F.~W.~J. Olver, A.~B. {Olde Daalhuis}, D.~W. Lozier, B.~I. Schneider,
                R.~F. Boisvert, C.~W. Clark, B.~R. Miller, B.~V. Saunders,
                H.~S. Cohl, and M.~A. McClain, eds."
}

\end{document}